\documentclass[12pt]{article}
\usepackage{amsmath,amssymb}
\usepackage[backend=bibtex,style=ieee]{biblatex}
\usepackage{authblk}
\usepackage{ragged2e}
\usepackage{graphicx}
\usepackage{float}
\usepackage{braket}
\usepackage{enumerate}
\usepackage{epigraph}
\usepackage{caption}
\usepackage[T1]{fontenc}
\usepackage{newpxtext}
\usepackage[bigdelims,vvarbb,varg]{newpxmath} 

\usepackage[dvipsnames]{xcolor}
\usepackage{todonotes}
\usepackage[linktocpage=true]{hyperref}
\usepackage[capitalise]{cleveref}
\usepackage{soul}

\hypersetup{
    colorlinks=true,         
    linkcolor=Blue,          
    citecolor=BlueViolet,        
    urlcolor=BlueViolet             
    }

\title{\bf Pure Shape Dynamics:\\ Relational General Relativity}

\author[a]{Pooya Farokhi\thanks{email: pooyafarokhi.p@gmail.com}}
\author[b]{Tim Koslowski\thanks{email:
tim.koslowski@thws.de}}
\author[c]{Pedro Naranjo\thanks{email: pnpfisica@gmail.com}}

\affil[a]{\normalsize \emph{University of Cologne, Department of Physics, Albertus-Magnus-Platz,} \break \emph{\normalsize 50923 Cologne, Germany}}
\affil[b]{\normalsize  \emph{Technical University of Applied Sciences W\"urzburg-Schweinfurt, Faculty of Applied Natural Sciences and Humanities, Ignaz-Sch\"on-Stra\ss{}e 11}, \break \emph{97421 Schweinfurt, Germany}}
\affil[c]{\normalsize \emph{Plaza Mayor, 4/1B, 09003 Burgos, Spain}}

\date{\normalsize \today}

\usepackage[a4paper,margin=1in]{geometry}

\newcommand\sbullet[1][.5]{\mathbin{\vcenter{\hbox{\scalebox{#1}{$\bullet$}}}}}

\begin{document}
\maketitle 

\begin{abstract}
\noindent We present a Pure Shape Dynamics (PSD) formulation of General Relativity (GR), which implements full relationalism by eliminating absolute scale and external time references from the fundamental description of gravity. Starting from the Arnowitt-Deser-Misner (ADM) formulation, we derive a decoupled dynamical system that governs the evolution of the spatial conformal geometry and relational matter degrees of freedom, while eliminating the total volume and York time as independent dynamical variables. This results in an autonomous subsystem describing an unparametrized trajectory in the conformal superspace of metric and matter configurations, with its evolution encoded in an equation of state that characterises the intrinsic geometric properties of the curve in shape space. We show that this equation of state is structurally analogous to the corresponding PSD description of the Newtonian $N$-body problem, reinforcing the fundamental similarity between gravity and relational particle dynamics. Our framework is applied to the homogeneous Bianchi IX cosmological model, demonstrating that the Janus point evolution through the Big Bang, as previously found in a symmetry-reduced setting, is a generic feature of the full inhomogeneous PSD description. This work establishes PSD as a fully scale- and reparametrization-invariant formulation of classical gravity and lays the foundation for addressing key open questions that are discussed at the end of the paper.

\end{abstract}
\newpage
\tableofcontents

\section{Introduction}\label{sec:intro}

Shape Dynamics (SD) (\cite{Barbour.2012}, \cite{Mercati.2018}) describes classical gravity as the evolution of spatial conformal geometry, in contrast to the description in terms of spacetime geometry, which underlies General Relativity (GR). This shift towards a relational description is forced upon any theory that intends to describe the universe as a whole, which includes by definition all reference processes and structures. This relational description can be constructed by considering so-called \emph{shape space} $\mathcal{S}\equiv \mathcal{Q}/\mathcal{G}$, where $\mathcal{Q}$ denotes the configuration space of the theory and $\mathcal{G}$ the symmetry group that leaves all observable ratios between configuration degrees of freedom and possible reference structures invariant. The transition from a description of dynamics on configuration space to a description on shape space implements \emph{spatial} relationalism.

A direct approach to impose spatial and \emph{temporal} relationalism invokes the so-called  \emph{strong Mach-Poincar\'e principle} (\cite{Barbour.2010}), which states that a point and a direction uniquely generate a curve in shape space. This however entails constant volume, thereby being unable to accommodate the observable expanding universe and structure formation, which thus renders the theory untenable (\cite{Anderson.2003})\footnote{The same drawback occurs in the particle model, where best-matching w.r.t dilatations implies vanishing of so-called dilatational momentum (\cite{Barbour.2003}).}.

To avoid this contradiction with observed structure formation, one can weaken the assumption and consider the \emph{weak Mach-Poincar\'e principle}: a point and a tangent vector uniquely determine a curve in $\mathcal{S}$. Interestingly, this theory is empirically viable \cite{Anderson.2005}. However, as emphasised in \cite{Vassallo.2022}, (Sec. 2.2), SD based on the weak Poicar\'e principle fails to describe classical gravity as a theory of the evolution of pure shape variables \emph{by themselves}, because the description requires an \emph{independent} degree of freedom, related to York time, to generate a parametrized curve in shape space. This is due to the fact that, in order for the theory to be empirically viable, the symmetry group is not the full set of conformal transformations, but its restriction to \emph{volume-preserving} conformal transformations, so the spatial volume, being the canonical conjugate of York time, is retained as a dynamical degree of freedom.     

The manifest implementation of temporal relationalism and the presence of a non-shape degree of freedom is what motivates the transition from SD to Pure Shape Dynamics (PSD) (\cite{Koslowski.2022} provides the fundamental ideas and applies them to the $N$-body system). The fundamental insight that enables the construction of PSD is the observation that \emph{temporal} relationalism implies that only ratios of change can be objectively quantified, e.g. as ratios of the change of state of a subsystem and the reference system that serves as the clock. This implies, geometrically, that the maximal objective description of the history of the universe is completely contained in the pure (i.e. unparametrized) curve in shape space. Hence, the objective dynamics of the universe is described by an \emph{equation of state of the geometry} of the pure curve in shape space.  This equation of state is by construction independent of the units used to describe the total spatial scale and the total duration. This implies the decoupling of the objective dynamics from the last dimensionful degree of freedom. The equation of state of the geometry of the curve in shape space can therefore be described in terms of the point is shape space, the tangent direction in shape space and two curvature degrees of freedom $\kappa$, $\epsilon$ that encode up to two dimensionless ratios that can be formed from the canonical shape degrees of freedom. Thus, while SD is a gauge theory of volume-preserving conformal transformations, PSD is obtained as a dynamically decoupeld subsystem of a gauge theory of unrestricted conformal transformations.

A very gratifying feature of PSD is the exhibition of the explicit structural similarity between the $N$-body system and dynamical geometry, as illustrated in \ref{subsec:equgravmatt}, \cref{eq:expliciteqs}, and \ref{subsec:comparison}, \cref{eq:particlePSD}.

The equivalence between SD and GR can be worked out using the linking theory (\cite{Gomes.2011, Gomes.2012a}) by gauge fixing the linking theory either in conformal gauge, which produces GR, or in ADM-gauge, which produces SD. The relation between SD and PSD is based on the identification of an autonomous dynamical subsystem within SD, which evolves the spatial conformal geometry by itself.

This paper is structured as follows. In Sec. \ref{sec:strategy} we will outline our approach: rather than using the original conformal techniques, which make use of the Constant Mean extrinsic Curvature (CMC) condition through the Lichnerowicz-York equation, we shall follow a more direct route by means of a suitable ``decoupling value'' $N_0(x)$ of the lapse, whereby the dynamics of the objective relational degrees of freedom decouple from the subjective conventional degrees of freedom. This is carried out in Sec. \ref{sec:decoupling}. The explicit dynamics is given in Sec. \ref{sec:shapeequ}. Firstly, in Sec. \ref{subsec:equgravmatt} we will provide the equation of state of the dynamical system gravity-matter, where the two crucial curvature degrees of freedom, $\kappa$ and $\epsilon$, are explicitly coupled to the shape and direction, which is key to effectively implementing full relationalism \emph{without} any non-shape degree of freedom. Secondly, Sec. \ref{subsec:comparison} provides an instructive illustration of the structural similarity between particles and dynamical geometry. To close this section, in Sec. \ref{sec:metricmeani} we shall take stock and have a look at the bigger picture: relationalism is clearly incompatible with geometry, conformal or otherwise, being a physical entity. The general SD programme, including PSD, still carries this conceptual flaw. Tentative ideas towards rectifying this are given. Finally, Sec. \ref{sec:cosmological} analyses the application of the general dynamical system to the cosmological realm: first, Sec. \ref{subsec:reduction} shows the associated symmetry reduction, and then the interesting case of Bianchi IX model is explicitly given in Sec. \ref{subsec:BianchiIX}. The paper closes with some open questions and future lines of research.  

\section{Strategy}\label{sec:strategy}

Local scale is a ratio between a geometric degree of freedom and a matter degree of freedom that acts as a local rod. Likewise, duration is a ratio of an amount of objective change of a process compared with the amount of a reference process that acts as a clock. It follows that a PSD description of gravity has to identify the autonomous subsystem of the dynamics that determines the ratios of the relative changes of local conformal geometry.

For the explicit construction of this autonomous subsystem, we start with the Arnowitt-Deser-Misner (ADM) description of gravity in terms of a spatial metric $g_{ab}(x)$, its canonically conjugate momentum density $p^{ab}(x)$, matter fields $\phi_A(x)$ and their canonically conjugate momentum densities $\sigma^A(x)$ (see, for instance, \cite{Gourgoulhon.2012} for an introduction, and \cite{Choquet.2009} for a mathematically rigorous treatment). Using the canonical Poisson bracket
\begin{equation}
  \begin{array}{rcl}
    \{f_1,f_2\}&:=&\int d^3x\left(\frac{\delta f_1}{\delta g_{ab}(x)}\frac{\delta f_2}{\delta p^{ab}(x)}-\frac{\delta f_1}{\delta p^{ab}(x)}\frac{\delta f_2}{\delta g_{ab}(x)}\right.\\
    && \left.\quad\quad\quad+\frac{\delta f_1}{\delta \phi_A(x)}\frac{\delta f_2}{\delta \sigma^A(x)}-\frac{\delta f_1}{\delta \sigma^A(x)}\frac{\delta f_2}{\delta \phi_A(x)}\right),
    \end{array}
\end{equation}
the canonical equations of motion $\dot f=\{f,H\}$ are generated by means of the ADM Hamiltonian:
\begin{equation}\label{eq:ADMcons}
  \begin{array}{ccl}
    H[N,\xi]&=&S[N]+H[\xi]\\
    S[N]&=&\int d^3x\,N\,\left(\frac{p^{ab}(g_{ac}g_{bd}-\frac{1}{2}g_{ab}g_{cd})p^{cd}}{\sqrt{g}}-\sqrt{g}(R[g]-2\Lambda-k\,H_{m}[g,\phi,\sigma])\right)\\
    H[\xi]&=&\int d^3x\,\left(p^{ab}\mathcal L_\xi g_{ab}+\sigma^A\mathcal L_\xi \phi_A\right),
  \end{array}
\end{equation}
with $N(x)$ being the so-called ``lapse'' scalar function, $\xi^a(x)$ being the so-called ``shift'' vector field, $\Lambda$ being the cosmological constant, $k$ being the gravitational coupling constant, $H_{m}[g,\phi,\sigma;x)$ being the scalar version of the matter Hamiltonian and $\mathcal L$ being the Lie derivative. The initial values for the canonical variables $(g_{ab},p^{ab},\phi_A,\sigma^A)$ have to satisfy the initial value constraints
\begin{equation}
  S[N]=0\quad\forall N(x),\quad\textrm{ and }\quad H[\xi]=0\quad\forall \xi^a(x).
\end{equation}

The starting point is the basic observation of SD that it is possible to ``gauge away'' the coupling between the evolution equations, on one hand, of the spatial conformal metric, parametrized e.g. by the density weight $-\frac 2 3$ unimodular metric $\rho_{ab}(x):=|g|^{-\frac 1 3}(x)g_{ab}(x)$, and its conjugate momentum, $\pi^{ab}(x)$, and, on the other hand, of the local volume element, parametrized e.g. by the weight 2 density $\omega_0(x)=\frac{|g|(x)}{v}$, with $v=\int d^3x \sqrt{|g|}$ being the spatial volume. This choice of variables separates the local scale from the total volume.

The fact that the evolution of the conformal metric decouples from the local scale degree of freedom was first observed in the CMC gauge-fixed version of the ADM system, which can be gauge-unfixed to a system that possesses local spatial Weyl transformations as gauge symmetries. Unfortunately, the construction of the gauge theory of local conformal transformations requires the explicit solution of Hamiltonian constraints for the local conformal factor. This Lichnerowicz-York equation is practically impossible to solve explicitly. Therefore, we shall follow a direct strategy that allows us to derive explicit equations of motion that reduce the dynamical system to the conformal degrees of freedom: we will consider a ``decoupling value'' $N_0(x)$ for the lapse that ensures that the equations of motion of the conformal metric $\rho_{ab}(x)$ and the matter degrees of freedom $\phi_A(x)$ decouple from the local volume element $\omega_0(x)$ and the trace of the metric momenta $p(x):=g_{ab}(x)p^{ab}(x)$. To do this we observe the following:
\begin{enumerate}
  \item For any value $\xi^a(x)$ of the shift vector field, one can find a lapse $N_0(x)$ that ensures that the evolution of the volume element be spatially constant, i.e.,
  \begin{equation}
    \frac{d \,\ln(\sqrt{|g|}(x))}{d\,t}=c.  
  \end{equation}   
    \item After deriving the equations of motion, one can replace $\frac{p(x)}{\sqrt{|g|}(x)}$ with the solution of the Hamiltonian constraint $S(x)=0$. In general, this solution comes with a sign ambiguity, but can be fixed using the principle that the gravitational arrow of time points in the direction of expanding spatial volume.
\end{enumerate}

In this way, we obtain a dynamical system that involves only $\rho_{ab}(x)$, its conjugate momentum $\pi^{ab}(x)$, the matter degrees of freedom $\phi_A(x), \sigma^A(x)$, as well as the total volume $v$. Remarkably enough, this decoupled dynamical system is, on the one hand, equivalent to the ADM formulation of GR and, on the other hand, possesses the same structure as the finite-dimensional PSD dynamical systems, which can be described as equations of state of the geometric properties of an unparametrized curve in the associated shape space. The decoupled ADM system can thus be turned into the equation of state of a curve in the relevant shape space, known as \emph{conformal superspace}, $\mathsf{S}=\mathsf{Superspace}/\text{conformal transformations}$, where $\mathsf{Superspace}=\mathsf{Riem}/\mathsf{Diff}(3)$, with 
$\mathsf{Riem}$ being the set of Riemannian $3$-geometries and $\mathsf{Diff}(3)$ the group of spatial diffeomorphisms. For the purpose of describing the evolution of gravity and matter, we enlarge conformal superspace to include the configuration degrees of the matter fields.

The important technical step that enables the construction of a temporally relational description of the equations of motion is the parametrization of the curve in shape space using an intrinsic parametrization condition, e.g. the arc-length $ds$ w.r.t. a metric on shape space. We should like to emphasise that this metric on shape space does not introduce a scale, but builds a ``democratic average'' of the instantaneous change of shape, that can be used to define an objective notion of duration. It follows, then, from the fact that $\rho_{ab}$ is dimensionless and scale-invariant that the democratically measured relative changes $\frac{d\,\rho_{ab}(x)}{ds}$ and $\frac{d\,\hat \phi_A(x)}{ds}$, with $\hat \phi_A$ being the objective ratios of matter degrees of freedom $\phi_A$, are also scale and reparametrization invariant. Thus, the RHS of these equations of motion are necessarily expressible in terms of scale- and reparametrization- invariant degrees of freedom. The same argument applies to the equations of motion of the scale- and reparametrization-invariant variables that appear on the RHS. This process can thus be iterated until the equations of motion close, i.e. until the scale- and reparametrization-invariant equations of motion of all variables that appear on the RHS are derived in this manner. This system of scale- and reparametrization-invariant equations of motion (cf. \cref{eq:expliciteqs}) depends by construction only on the local geometry of the curve in shape space and can thus be expressed as an equation of state of the geometry of said curve in shape space.

In the next section we shall explicitly perform the decoupling of the physical degrees of freedom, which will allow us to derive, in Sec. \ref{sec:shapeequ}, the equation of state of the curve in shape space.

\section{Decoupling of the Relational Degrees of Freedom}\label{sec:decoupling}

To prepare the construction of the decoupling of the physical degrees of freedom, it is useful to introduce a redundant canonical transformation that separates the conformal metric $\rho_{ab}(x)$ from the local volume degree of freedom $\omega_0(x)$ and the spatial volume $v$. This can be obtained by means of the following generating functional:
\begin{equation}
  F=\int d^3x\left(\pi^{ab}(x)|g|^{-\frac 1 3}(x)g_{ab}(x)+\pi_0(x)\frac{|g|(x)}{\left(\int d^3x \sqrt{|g|}\right)^2}+\tau\,\sqrt{|g|}(x)\right), 
\end{equation}
which allows us to readily derive the relation between the configuration variables thus:
\begin{equation}\label{canonical}
  \begin{array}{ccl}
    \rho_{ab}(x)&=&\frac{\delta\,F}{\delta\,\pi^{ab}(x)}=|g|^{-\frac 1 3}(x)g_{ab}(x)\\
    \omega_0(x)&=&\frac{\delta\,F}{\delta\,\pi_0(x)}=\frac{|g|(x)}{v^2}\\
    v&=&\frac{\delta\,F}{\delta\,\tau}=\int d^3x\sqrt{|g|}(x).
  \end{array}
\end{equation}

The redundancy of this canonical transformation leads to reducibility conditions, in particular the unimodular and unit volume constraints,
\begin{equation}
  |\rho|(x)=1\textrm{ (unimodular) and }\int d^3x\sqrt{\omega_0}(x)=1\textrm{ (unit volume),}
\end{equation}
and further conditions that we subsequently derive as invertibility conditions on the configuration variables. Calculating the relation between the momentum variables yields at once:
\begin{equation}
 \begin{array}{l}
  p^{ab}(x)=v^{-\frac 2 3}\left(\omega_0^{-\frac 1 3}(x)\left(\delta^a_c\delta^b_d-\frac 1 3\rho^{ab}(x)\rho_{cd}(x)\right)\pi^{cd}(x)+\omega_0^{\frac 2 3}(x)\rho^{ab}(x)\pi_0(x)\right.\\
  \quad\quad\quad\quad\quad\quad\left.-\omega_0^{\frac 1 6}(x)\rho^{ab}(x) \,\omega_0 \sbullet \pi_0\right)
  +\frac 1 2 v^{\frac 1 3} \omega_0^{\frac 1 6}(x)\rho^{ab}(x)\tau, 
 \end{array}
\end{equation}
where we have used the shorthand  $\omega_0\sbullet\pi_0 :=\int d^3x\,\omega_0(x)\,\pi_0(x)$ and where $\rho^{ab}(x)$ refers to the inverse conformal metric $|g|^{\frac 1 3}(x)g^{ab}(x)$. The inversion of these relations is singular. First, notice that due to the occurrence of the trace-free projection $\left(\delta^a_c\delta^b_d-\frac 1 3\rho^{ab}(x)\rho_{cd}(x)\right)$, we cannot solve for the trace of $\pi^{ab}(x)$. Thus, in order for our analysis to be independent of this choice, let us impose the trace-less constraint:
\begin{equation}
  \rho_{ab}(x)\pi^{ab}(x)=0,
\end{equation} 
which implies $\left(\delta^a_c\delta^b_d-\frac 1 3\rho^{ab}(x)\rho_{cd}(x)\right)\pi^{cd}(x)=\pi^{ab}(x)$, which we will use in the following. Moreover, we observe from 
\begin{equation}
  \int d^3x g_{ab}(x)\pi^{ab}(x)=\frac{3}{2}v\,\tau
\end{equation}
that one cannot solve for $\omega_0 \sbullet \pi_0$. We thus impose the zero-mean constraint:
\begin{equation}
  \omega_0 \sbullet \pi_0=0
\end{equation}
on $\pi_0(x)$. In summary, from \cref{canonical}, we have arrived at the following canonical transformation:
\begin{equation}\label{canonical2}
 \begin{array}{ccl}
  p^{ab}(x)&=&v^{-\frac 2 3}\left(\omega_0^{-\frac 1 3}(x)\pi^{ab}(x)+\omega_0^{\frac 2 3}(x)\rho^{ab}(x)\pi_0(x)\right)+\frac 1 2 v^{\frac 1 3} \omega_0^{\frac 1 6}(x)\rho^{ab}(x)\tau\\
  g_{ab}(x)&=&v^{\frac 2 3}\,\omega_0^{\frac 1 3}(x)\,\rho_{ab}(x), 
 \end{array}
\end{equation}

and the set of conjugate variables $\left\{(\rho_{ab}(x),\pi^{ab}(x)), (\omega_0(x), \pi_0(x)), (v,\tau)\right\}$, together with the reducibility conditions
\begin{equation}
  \begin{array}{rclcrcl}
    |\rho|(x)&=&1,&\quad&\int d^3x\sqrt{\omega_0}(x)&=&1,\\
    \rho_{ab}(x)\pi^{ab}(x)&=&0,&\quad&\int d^3x\, \omega_0(x)\pi_0(x)&=&0.
  \end{array}
\end{equation}
The canonical transformation \cref{canonical2} allows us to express $S[N]$ and $H[\xi]$ as:
\begin{equation} \label{eq:newconstraints}
 \begin{array}{l}
  S[N]=\int d^3x\,N(x)\,\left(\frac{\pi^{ab}\rho_{ac}\rho_{bd}\pi^{cd}}{v\,\sqrt{\omega_0}} -\frac{p^2}{6\,v\,\sqrt{\omega_0}}-v^{\frac 1 3}\sqrt{\omega_0}R[\omega_0^{\frac 1 3}\rho_{ab}]+2\,\Lambda\,v\,\sqrt{\omega_0}\right. \\ \quad\quad\quad\quad\quad\quad\quad\quad\quad\quad\quad\left.+v\,\sqrt{\omega_0}H_{m}[v^{\frac 2 3}\omega_0^{\frac 1 3}\rho_{ab},\phi_A,\sigma^A]\right),\\
  H[\xi]=\int d^3x\left(\pi^{ab}\mathcal L_\xi \rho_{ab}+\omega_0\left(\pi_0+\frac{v\,\tau}{2\sqrt{\omega_0}}\right)\rho^{ab}\mathcal L_\xi\rho_{ab}+\sigma^A\mathcal L_\xi \phi_A\right),
 \end{array}
\end{equation}
where we have used $0=\int d^3x \mathcal L_\xi \sqrt{\omega_0}(x)=\frac 1 2 \int d^3x \,\omega_0^{-\frac 1 2}(x)\mathcal L_\xi \omega_0(x)$ and the shorthand $p(x) = 3\left(\omega_0(x)\pi_0(x)+\frac 1 2\sqrt{\omega_0}(x) v\tau\right)$.

The shape component of the kinetic term in the expression for $S[N]$ in \cref{eq:newconstraints} can be written in terms of the ``shape space supermetric''

\begin{equation}
G^S_{abcd} := \rho_{ac}\rho_{bd} + \rho_{ad} \rho_{bc}
\end{equation}
such that the kinetic term becomes
\begin{equation}
\pi^{ab}\rho_{ac}\rho_{bd}\pi^{cd} = \frac{1}{2} G^S_{abcd} \pi^{ab} \pi^{cd}.
\end{equation}

To decouple the evolution of the relational degrees of freedom from the local volume element, we impose the decoupling condition

\begin{equation}\label{eq:decouplingCond}
  c=\frac{d}{d\,t}\ln\left(\sqrt{|g|}(x)\right),
\end{equation}
where $c$ is a dimensionless constant that is chosen once and for all. The constancy of $c$ ensures that $\omega_0(x)$ have no dynamics. Using 
\begin{equation*}
  \frac{d \ln\sqrt{|g|}}{d\,t}=\frac{1}{2}g^{ab}(x)\dot{g}_{ab}(x)=\frac 1 2 \left(g^{ab}(x)\mathcal L_\xi g_{ab}(x)-N(x)\frac{p(x)}{\sqrt{|g|}(x)}\right)  
\end{equation*}
and solving the scalar constraint $\frac{S(x)}{\sqrt{|g|}(x)}=0$ for 

\begin{equation*}
 \left(\frac{p(x)}{\sqrt{|g|}(x)}\right)^2=6\left[\frac{\pi^{ab}\rho_{ac}\rho_{bd}\pi^{cd}}{v^2\omega_0}-v^{-\frac 2 3}R[\omega_0^{\frac 1 3}\rho_{ab}]+2\Lambda+H_{m}\right],    
\end{equation*}
we find that condition (\ref{eq:decouplingCond}) is satisfied when the lapse takes the decoupling value
\begin{equation}\label{eq:lapse}
N_0(x)=\pm\frac{\left|g^{ab}\mathcal L_\xi g_{ab}-2c\right|}{\sqrt{6\,\left|\frac{\pi^{ab}\rho_{ac}\rho_{bd}\pi^{cd}}{v^2\omega_0}-v^{-\frac 2 3}R[\omega_0^{\frac 1 3}\rho_{ab}]+2\Lambda+H_{matt.}\right|}},
\end{equation}
where the sign depends on whether we are considering a contracting or expanding branch. The denominator vanishes \emph{only} in the ``maximal CMC'' slice, which is the point of minimal structure formation and therefore of minimal records of the past in an expanding universe. Using this fact, the gravitational arrow of time suggests that the experienced direction of time points into the direction of spatial expansion, so we choose the positive sign from now on.

\section{Shape Equations of State}\label{sec:shapeequ}

\subsection{PSD equations of gravity and matter}\label{subsec:equgravmatt}

\Cref{canonical,eq:newconstraints} readily yield the decoupled equations of motion (to avoid cumbersome notation, we set the shift vector field $\xi^a(x)\equiv 0$):

\begin{equation}\label{eq:decoupledeqs}
 \begin{array}{ccl}
   d\rho_{ab}(x)&=&\int d^3y\,N_0(y)\,\frac{\delta S(y)}{\delta \pi^{ab}(x)},\\
   d\pi^{ab}(x)&=& -\int d^3y\,N_0(y)\,\frac{\delta\,S(y)}{\delta \rho_{ab}(x)},\\
   d\phi^A (x)  &=& \int d^3y\,N_0(y)\,\frac{\delta\,S(y)}{\delta \sigma_A (x)},\\
   d\sigma_A (x)  &=& -\int d^3y\,N_0(y)\,\frac{\delta\,S(y)}{\delta \phi^A (x)},\\
   d v&=&v c.
 \end{array}
\end{equation} 

These equations are not yet in the dimensionless form that we desire, since $\pi^{ab}(x)$, $\sigma _A (x)$, $v$ and $\Lambda$ are dimensionful quantities\footnote{For the sake of argument and without loss of generality, we can assume that all coupling constants in the matter sector have been rendered dimensionless by multiplying them with appropriate powers of the cosmological constant.}. Let us next define
\begin{equation}
\Sigma:=\int\,d^3x\sqrt{G^S_{abcd} \pi^{ab}\pi^{cd}+T_\text{kin}(\sigma)},
\end{equation}
where $T_\text{kin}(\sigma)$ is a density weight 2 quadratic form of the matter momenta $\sigma^A(x)$, that essentially denotes the matter kinetic energy density. This way, we can define dimensionless relational momenta as
\begin{equation}
  \hat{\pi}^{ab}(x):=\frac{\pi^{ab}(x)}{\Sigma},\quad \hat \sigma^A(x):=\frac{\sigma^A(x)}{\Sigma},
\end{equation} 
as well as the dimensionless ratios
\begin{equation}
\label{kappa_epsilon}
  \kappa:=\Sigma^2\,v^{-\frac 4 3},\quad \epsilon:=\frac{\Lambda\,v^2}{\Sigma^2},
\end{equation}
which absorb the remaining dimensional degrees of freedom $v$, $\Sigma$ and $\Lambda$. With these definitions in place, it follows that we can express the equation of state of the geometry of the curve in shape space in terms of:
\begin{enumerate}
 \item The point in shape space, parametrized by relational degrees of freedom $\rho_{ab}(x)$ and $\phi_A(x)$.
 \item The tangent direction, parametrized by the normalized momenta $\hat \pi^{ab}(x)$, $\hat \sigma^A(x)$.
 \item Two degrees of freedom directly related to the curvature degrees of the curve in shape space, parametrized by the dimensionless ratios $\kappa$, $\epsilon$.
\end{enumerate} 

Before calculating the equation of state, let us make three observations:
\begin{enumerate}
 \item One can replace the $\pi^{ab}(x) = \Sigma \hat \pi^{ab}(x)$ in $S(y)$ when calculating the derivative w.r.t. $\rho_{cd}(x)$ without changing the equations of motion.
 \item One can make the solution for the lapse function in~\cref{eq:lapse} scale-invariant. A quick observation reveals that 
\begin{equation}
\hat{N}_0(x) = \frac{\Sigma}{v} N_0 (x) = \frac{2c}{\sqrt{6\,\left|\frac{\hat\pi^{ab}\rho_{ac}\rho_{bd}\hat\pi^{cd}}{\omega_0}-\frac{1}{\kappa}R[\omega_0^{\frac 1 3}\rho_{ab}]+2\epsilon+\kappa^{-\frac{3}{2}}H_{matt.}\right|}}
\end{equation}
is scale-invariant, where $\hat{H}_m[\rho_{ab},\phi_A,\hat\sigma^A,\kappa,\epsilon] := \Sigma H_m$ is the scale-invariant Hamiltonian for matter fields.
\item The equation for $\Sigma$ reads\footnote{For the sake of simplicity, in the following we will omit $T_\text{kin}(\sigma)$, since it unnecessarily convolutes the equations without adding any conceptual benefit.}
\begin{equation}\label{eq:eqsigma}
 \begin{array}{ccl}
   d\Sigma/ \Sigma &=& \int d^3x \frac{\hat{N}_0 G^S_{abcd} \hat{\pi}^{cd}}{\sqrt{G^S_{abcd} \hat{\pi}^{ab} \hat{\pi}^{cd}}} \left(\frac{\sqrt{\omega_0}}{\kappa} \frac{\delta R}{\delta\rho_{ab}} - \sqrt{\omega_0} \kappa^{-\frac{3}{2}}  \frac{\delta \hat{H}_m}{\delta\rho_{ab}}\right).
 \end{array}
\end{equation}
\end{enumerate}

With these preparations at our disposal, we are able to derive the explicit equations of motion that follow from~\cref{eq:decoupledeqs} in terms of only scale-invariant variables, namely the equation of state of the curve in shape space, which reads thus:

\begin{equation}
\label{eq:expliciteqs}
\begin{array}{ccl}
d\rho_{ab}(x)&=& \frac{\hat{N}_0 (x)}{\sqrt{\omega_0}} (x) \, G^S_{abcd} \hat{\pi}^{cd} (x), \\
   d{\hat{\pi}}^{ab}(x)&=& \int d^3y \, \frac{\delta \hat{\pi}^{ab}(x)}{\delta \rho_{cd}(y)} d\rho_{cd} (y) - \int d^3y \, \hat{N}_0(y) \frac{\delta \hat{\pi}^{ab}(x)}{\delta \pi^{cd}(y)/\Sigma} \left(\frac{\hat{\pi}^{ce} \hat{\pi}^{df} \rho_{ef}}{\sqrt{\omega_0}} -\frac{\sqrt{\omega_0}}{\kappa} \frac{\delta R}{\delta\rho_{cd}} + \sqrt{\omega_0} \kappa^{-\frac{3}{2}} \frac{\delta \hat{H}_m}{\delta\rho_{cd}} \right)(y),\\
   d{\phi}^A(x)&= & \hat{N}_0 (x) \left(\sqrt{\omega_0} \kappa^{-\frac{3}{2}} \frac{\delta \hat{H}_m}{\delta \hat\sigma_A} (x) \right),\\
   d{\hat\sigma}_A(x)&= & -\hat{N}_0 (x) \left(\sqrt{\omega_0} \kappa^{-\frac{3}{2}} \frac{\delta \hat{H}_m}{\delta \phi^A}(x) \right), \\
   d\kappa &= & 2\kappa \int d^3x \frac{\hat{N}_0 G^S_{abcd} \hat{\pi}^{cd}}{\sqrt{G^S_{abcd} \hat{\pi}^{ab} \hat{\pi}^{cd}}} \left(\frac{\sqrt{\omega_0}}{\kappa} \frac{\delta R}{\delta\rho_{ab}} - \sqrt{\omega_0} \kappa^{-\frac{3}{2}} \frac{\delta \hat{H}_m}{\delta\rho_{ab}}\right) -\frac{4}{3} \kappa c,\\
   d\epsilon &=& - 2 \epsilon \int d^3x\,  \frac{\hat{N}_0 G^S_{abcd} \hat{\pi}^{cd}}{\sqrt{G^S_{abcd} \hat{\pi}^{ab} \hat{\pi}^{cd}}} \left(\frac{\sqrt{\omega_0}}{\kappa} \frac{\delta R}{\delta\rho_{ab}} - \sqrt{\omega_0} \kappa^{-\frac{3}{2}} \frac{\delta \hat{H}_m}{\delta\rho_{ab}}\right) + 2c \epsilon.
\end{array}
\end{equation}

Let us pause now to analyse the interpretation of the dynamical system  \cref{eq:expliciteqs}. As in the particle model, PSD describes dynamical trajectories ``purely'' in shape space, the space of (diffeomorphism classes of) $\rho_{a b}$'s in this case. As outlined above, this description is in terms of the shape variable itself, the direction of its change, and further geometrical quantities of the trajectory, which can be expressed in terms of the curvature and jerk of the curve in shape space.

To see this, let us first consider the directions (for the sake of simplicity, let us only take the $\hat{\pi}^{a b}$'s). Judging from the RHS of the equation for $d \rho_{a b}$, a one-to-one function of $\hat{\pi}^{a b}$'s can define the "angles" of directions. Given the constraint $\int d^3x \, \hat\pi^{a b} \hat\pi^{cd} G^S_{abcd}=1$, such function has to be homogeneous of degree 0 in $\hat{\pi}^{a b}$'s. Let us call the angle variable of the changes at a given shape $\psi^{ab}$. There is a bijective relation between $\hat{\pi}^{a b}$ and $\psi^{a b}$, so $d \rho_{a b}$, $d \psi^{a b}$, $d \kappa, d \epsilon, d \phi, d \sigma$ can all be written in terms of $\rho_{ab}(x), \psi^{a b}(x)$ instead of momenta.

As for other variables, one can use a reference metric in shape space and derive the curvature and jerk of the curve in shape space and express these geometric quantities through $d \psi^{a b}$, \ldots As an explicit example, let us assume pure dynamical geometry ($H_m=0$) with $\Lambda=0$,  for simplicity. The only additional variable with no apparent shape interpretation is $\kappa$, which appears in the form $\frac{1}{\kappa} \delta R$ in the equations. Therefore, $\kappa$ is related to how much the curve ``deviates'', at the second order, from the shape trajectory of a free system with no potential $-R$ appearing, that is,

\begin{equation}
S'{[N]}=\int d^3x \, N(x) \left(T_S[\rho_{ab}, \pi^{ab},v]+T(v,p)\right).
\end{equation}

To quantify it, assuming the same shape and direction, let $d \psi^{ab}$ be the next-order change in the original model, and $d \psi'^{a b}$ the one corresponding to the free model $S'$. Then let

\begin{equation}
C \equiv G_{a b c d}^S\left(d \psi^{a b^{\prime}}-d \psi^{a b}\right)\left(d \psi^{\prime c d}-d \psi^{c d}\right)=\frac{1}{\kappa^2} \int_{\Sigma} d^3x \, \omega_0(x) \hat{N}_0^2(x) G_{a b c d}^S \frac{\delta R}{\delta\rho_{ab}} \frac{\delta R}{\delta\rho_{cd}}
\end{equation}
define the curvature of the trajectory. 

Therefore, $\kappa$ measures the ``radius of curvature" of shape trajectories w.r.t those of the free model. Remarkably enough, the whole dynamical system \cref{eq:expliciteqs} can be fully cast in terms of $\rho_{a b}, \psi^{a b}$, and $C$, which all have clear interpretations in shape space. This interpretative approach can be implemented regardless of the complexity of the model. 

\subsection{Comparison with the $N$-body Newtonian system}\label{subsec:comparison}

The same strategy as spelled out in Sec.~\ref{sec:strategy} has been applied to the Newtonian $N$-body system, which results in the particle PSD description of the model \cite{Koslowski.2022}. Interestingly, due to the same conceptual motivations as the particle case and structurally identical set of principles and strategies, the final systems of equations of state in both cases bear a marked similarity.

In the particle model, the finite-dimensional compact shape space $\mathcal{S}$ is equipped with coordinates $q^a$ whose conjugate variables in the phase space representation are $p_b$. Total scale, defined as the moment of inertia of the system, $R$, gives place to the scale-invariant quantity 

 \begin{equation}
\label{kappa}
   \kappa = p^2/R,   
 \end{equation}
which is the counterpart of \cref{kappa_epsilon}. The system closes with the following equation of state:
\begin{equation}\label{eq:particlePSD}
\begin{array}{ccl}
  d q^a &=& u^a(\phi),\\
  d \phi_A &=& \frac{\partial \Phi_A}{\partial q^a}u^a(\phi)-\frac{\partial \Phi_A}{\partial u^a} \left(\frac 1 2 g^{bc}_{,a}(q)u_b(\phi)u_c(\phi)
  +\frac 1 \kappa V_{,a}(q)\right),\\
  d \kappa &=& -2u^a(\phi)V_{,a}(q)\mp\kappa\sqrt{-\left(1+\frac{2\,V(q)}{\kappa}\right)}.\\
 \end{array}
\end{equation}
Note the parallel with \cref{eq:expliciteqs}. Firstly, given we can write
$
\hat{\pi}^{ac} \hat{\pi}^{bd} \rho_{cd} = \frac{1}{2} \frac{\delta G^S_{cdef}}{\delta\rho_{ab}} \hat{\pi}^{cd} \hat{\pi}^{ef},
$
the expression in the RHS of the equation for $d\hat{\pi}^{ab}$ takes on the same structure as the one for $d\phi_A$ in \cref{eq:particlePSD} once we identify $-R[\omega_0^{\frac 1 3}\rho_{ab}]$ as the potential, along with the $\rho_{ab}$-dependent part of the matter Hamiltonian.

Beyond the almost identical structure of dynamics, two main differences must be highlighted: 1. In \cref{eq:expliciteqs} there are distinct, but coupled, sets of shape variables for the metric and the material fields, due to the ``double ontology'' of GR, which has no counterpart in the particle model (see Sec.~\ref{sec:metricmeani} below). 2. Whereas the first terms in the RHS of the equations for $\kappa$ and $\epsilon$ in \cref{eq:expliciteqs} are on an equal footing with the one for $\kappa$ in \cref{eq:particlePSD}, with all of which resulting from the equation for $\Sigma$ (or $p$ in the particle model), the counterpart of the term $\mp\kappa\sqrt{-\left(1+\frac{2\,V(q)}{\kappa}\right)}$, which signifies Newtonian expansion/contraction of the universe, respectively, is missing in the field model. This is because in the particle model, once we transition to scale-invariant dimensionless variables and $\kappa$ is defined similarly, one has to solve the Hamiltonian constraint for the variable conjugate to total scale (dilatational momentum) and thereby close the dynamical system. In the current model, however, our strategy has been to decouple the local scale variables $\{\omega_0, \pi_0\}$ by fixing the ``local'' parametrization through $N(x)$. This already fixes the dynamics of volume. On the other hand, the Hamiltonian constraint has been solved for (the decoupling value of) the lapse function, $N_0(x)$. As a result, the local nature of scale in the current gravitational model implies these different implementations of our strategy for the construction of the PSD description of the two models.

\subsection{Interlude: the meaning of $\rho_{ab}(x)$}\label{sec:metricmeani}
The aspiration of the SD programme has long been the construction of a dynamical theory of geometry in which the physically relevant degrees of freedom all exist in conformal superspace (defined in Sec. \ref{sec:strategy}) (\cite{Anderson.2003, Barbour.18092010}). This has been realized in the current PSD formalism in Eq.~\eqref{eq:expliciteqs} through decoupling the scale and restricting the dynamically relevant degrees of freedom to the metric shapes $\rho_{ab}$. The original motivation for such reduction is grounded on the interpretation of the spatial metric function as determining ``physical distance''. If $g_{ab} dx^a dx^b$ measures physical length, in light of the conceptual principles of SD, which argue that only the ratios of distances are observable and thus should be physically relevant, one has to identify the scale-invariant ``distance-determining'' part of $g_{ab}$. Due to coordinate freedom, invariance under diffeomorphisms already picks out the relevant distance information contained in $g_{ab}$. Once we peel away the scale, the result will be the shape content of the metric, $\rho_{ab}$, as laid out above. 

However, a somewhat more fundamental problem arises as to the physical meaning of metric in the first place.  It is widespread received wisdom that Einstein's gravity, by elevating the (suitable generalisation of the) fixed background Euclidean geometry of Newtonian physics to a dynamical structure, suitably coupled to material fields, is a hallmark of theoretical physics. There are, however, legitimate reasons for criticizing the ``double ontology'' of GR, which treats matter and geometry degrees of freedom as two sides of the same coin, in particular in a completely relational framework. After all, there are two conceptually disparate geometries in this setup, the first being the explicit geometry that is treated as the gravitational field, while the second is the directly observable geometry that is deduced from the evolution of material reference systems and processes, the so-called clocks and rods. While the conceptual difference remains, these two geometries happen to coincide when the usual properties of matter fields hold (locality, standard kinetic term, minimal coupling, etc.), as has been shown in \cite{Gomes.2012b}.

The motivation for the PSD framework is the identification of the minimal autonomous subsystem that is sufficient to describe all physical experiences. It can be reasonably argued that any physical experience supervenes on matter configurations for it is not the case that the formal geometry $\rho_{ab}$ is equipped with observable physical content; rather, it is the material sector that comes endowed with observable geometrical structure. If these arguments are accepted as part of the foundations of a relational setup, then one is forced to start the construction of the autonomous dynamical subsystem with the relational matter degrees of freedom $\phi_A$ and describe the equation of state of the geometry of the curve in the matter shape space. This description has to be worked out on a case by case basis, depending on the precise matter field content, but it is generally true that the equation of state of the geometry on matter shape space will depend on local curvature degrees of freedom that encode the interaction of matter fields with spatial conformal geometry, i.e. the degrees of freedom described by the diffeomorphism classes of $\rho_{ab}$, since the shape of the matter configuration is e.g. affected by gravitational waves passing through a matter shape.

This leads us to declare that the true physical PSD model of gravity should permit only shapes inherent to material fields, and dispense with an independent geometry altogether, such as $\rho_{ab}$, as a fundamental physical degree of freedom. Nevertheless, in general, when enough non-vanishing matter field components are available, one will be able to invert these for the local spatial conformal geometry. It is therefore legitimate to consider the spatial conformal geometry as an effective part of the dynamical degrees of freedom (i.e. elements of shape phase space) of any PSD description, even when adopting a ``primacy'' of matter degrees of freedom.

\section{The Cosmological Model}\label{sec:cosmological}

As illustration of the general dynamical system \cref{eq:expliciteqs}, in this section we will analyse the quiescent Bianchi IX model, whose PSD description was worked out in \cite{Koslowski.2018}, where it was shown that the PSD description retains hyperbolicity through the Big Bang, thereby unambiguously connecting two Big Bang universes at this so-called Janus point. The purpose of the present section is to suggest that this result most likely does not depend on the symmetry reduction, but is expected to be a property of the dynamical system of full inhomogeneous PSD when evaluated at a homogeneous cosmological model (for further reference on homogeneous cosmology, see, e.g. \cite{Jha.2023} for a nice account within GR, and \cite{Mercati.2018} for a pedagogical rendition within SD). We start by carrying out the corresponding symmetry reduction in Sec. \ref{subsec:reduction}, and then proceed in Sec. \ref{subsec:BianchiIX} with the explicit dynamics of this cosmological model.  

\subsection{Symmetry reduction and variables}\label{subsec:reduction}

As is well-known, the Bianchi IX cosmological model is defined by homogeneous geometry on the 3-sphere with global degrees of freedom for the orientation. We will follow \cite{Koslowski.2018,Mercati.2018}. Using coordinates $(r, \theta, \phi)$ which satisfy $$0 \leq r \leq \pi, 0 \leq \theta \leq \pi, 0 \leq \phi \leq 2\pi,$$ we have the following translation generating one-forms $(\sigma^{i} \equiv d x^i)$:

\begin{equation}
\begin{array}{ccl}
\sigma^1 &=&\sin r d \theta-\cos r \sin \theta d \phi, \\
\sigma^2 &=&\cos r d \theta+\sin r \sin \theta d \phi, \\
\sigma^3 &=&-d r-\cos \theta d\phi .
 \end{array}
\end{equation}

We will use these three one-forms and their dual vectors, $\chi_i$'s, satisfying $\sigma^j\left(\chi_i\right)=\sigma_a^j \chi_i^a=\delta_i^j$,

\begin{equation}
\begin{array}{ccl}
\chi_1&=&\cos r \cot \theta \partial_r+\sin r \partial_\phi-\cos r \csc \theta \partial_\phi, \\
\chi_2&=&-\sin r \cot \theta \partial_r+\cos r \partial_\theta+\sin r \csc \theta \partial_\phi, \\
\chi_3&=&-\partial_r,
\end{array}
\end{equation}
as the triad variables. Therefore, the metric is written as

\begin{equation}
g_{a b}=q_{i j} \sigma_a^i \sigma_b^j\,, 
\end{equation}
where
$q_{i j}$ is a constant matrix, serving as the ``global'' metrical degrees of freedom of the homogeneous universe. Similarly,

\begin{equation}
p^{a b}=k^{i j} \chi_i^a \chi_j^b \, \frac{\left|\sigma^1 \wedge \sigma^2 \wedge \sigma^3\right|}{4 \pi^2},
\end{equation}
with constant $k^{i j}$. The factor $\left|\sigma^1 \wedge \sigma^2 \wedge \sigma^3\right| = | \sin \theta |$ is to ensure the correct weight of $p^{a b}$, and $\frac{1}{4 \pi^2}$ makes $k^{i j}$ conjugate to $q_{i j}$, which is the case, because the symplectic form reads:

\begin{equation}
\theta=\int d^3x \, p^{a b} g_{a b}=\int d^3x \, \frac{\sin \theta}{4 \pi^2} \delta k^{i j} \delta q_{i j}=k^{i j} \delta q_{i j} .
\end{equation}

We are now in the position to transition to shape variables by decomposing the global variables and decoupling scale. Direct calculation shows

\begin{equation}
\left|g_{ab}\right|=\left|q_{i j} \sigma_a^i \sigma_b^j\right|=\sin^2 \theta\left|q_{i j}\right|, 
\end{equation}
and therefore,
\begin{equation}
v = 4\pi^2 \sqrt{|q_{ij}|}, \quad \omega_0=\frac{|\sin^2 \theta|}{16\pi^4}.
\end{equation}
We define the shape variable and its conjugate momentum as

\begin{equation}
\rho_{ab} = \theta_{ij} \hat\sigma_a^i \hat\sigma_b^j  \quad
\pi^{a b}=\eta^{i j} \hat{\chi}_i^a \hat{\chi}_j^b,
\end{equation}
where the triads have been rescaled as $\hat\sigma^j = \omega_0^{-\frac{1}{6}} \sigma^i$ and $\hat\chi_i =  \omega_0^{\frac{1}{6}} \chi_i$ to match the correct weight of the shape conjugate variables. It immediately follows that

\begin{equation}
\theta_{ij} = \frac{q_{ij}}{v^{\frac{2}{3}}}\,,
\end{equation}
and therefore, $|\theta| = (16\pi^4)^{-1}$.
As for the relation between $\eta^{i j}$ and $k^{ij}$,

\begin{equation}
p^{ab}=\pi^{ab}|g|^{-\frac{1}{3}}+\frac{p}{3} g^{a b}, 
\end{equation}
which results in

\begin{equation}
\eta^{i j}=k^{i j} v^{\frac{2}{3}} -\frac{k}{3} g^{i j} v^{\frac{2}{3}}, 
\end{equation}
with $k=q_{i j} k^{i j}$ and $q^{i j}=\left(q\right)_{i j}^{-1}$.
The trace of $k^{ij}$ reads
\begin{equation}
k=3 \pi_0 \sqrt{\omega_0}+\frac{3 v\tau}{2},
\end{equation}
which, considering the choice $\pi_0 \sbullet \omega_0 = 0$, results in $\pi_0 = 0$ in the homogeneous case, leading to $k = \frac{3}{2} v\tau$.

Ultimately, in summary, we have the conjugate variables $\left(\theta_{i j}, \eta^{i j}\right)$ and $(v, \tau)$, satisfying the conditions

\begin{equation}\label{eq:cosmconstraints1}
|\theta|=(16\pi^4)^{-1},\quad \eta = \theta_{ij} \eta^{i j}=0.
\end{equation}
We must note that being scalars, the conjugate matter variables, $(\phi_A$, $\sigma^A)$, must be constant in the homogeneous case.

\subsection{Dynamical equations of Bianchi IX PSD}\label{subsec:BianchiIX}

To derive the equation of state associated with the Bianchi IX cosmological model, we can follow two strategies. One is to simply reduce the general system \cref{eq:expliciteqs} to variables $\theta_{ij}, \hat{\eta}^{i j}, \kappa, \epsilon, \phi_A, \hat{\sigma}^A$, defined accordingly. The other strategy, which we will be following here, is to derive the dynamical Bianchi IX system from the dynamical constraints $S$ and $H$ rewritten in terms of these global variables. We have
\begin{equation} \label{eq:cosmconstraintS}
 \begin{array}{l}
 S[N]=\int d^3x\, N\left(\frac{\sqrt{\omega_0} \mu_{ijkl} \eta^{ij}\eta^{kl}}{2v}  - \frac{\sqrt{\omega_0} k^2}{6v} - v^{\frac{1}{3}} \sqrt{\omega_0} R[\omega_0^{\frac{1}{3}} \theta_{ij} \hat{\sigma}_a^i \hat{\sigma}_b^j] \right. \\
\quad\quad\quad\quad\quad\quad\quad\quad\quad\quad\quad \left. + 2\Lambda v \sqrt{\omega_0} +v \sqrt{\omega_0} H_m\right),
 \end{array}
\end{equation}
where $\mu_{ijkl}$ is the symmetry-reduced shape space kinematic metric 

\begin{equation}
\mu_{i j k l}=\theta_{i k} \theta_{j l}+\theta_{i l} \theta_{j k},
\end{equation}
The lapse function $N$ has to be constant, and $R$ is also a constant function of only $\theta_{ij}$, as expected on symmetry grounds. Thus,

\begin{equation}
S(N)=\frac{1}{2} \frac{\mu_{ijkl} \eta^{ij} \eta^{kl}}{v}-\frac{k^2}{6 v}-v^{\frac{1}{3}} R(\theta_{ij}) + 2 \Lambda v + v H_m(v, \phi, \sigma),
\end{equation}
where the scalar curvature is explicitly

\begin{equation}
R(\theta_{ij})= \operatorname{Tr}\theta^2-\frac{1}{2}\left(\operatorname{Tr}\theta\right)^2.
\end{equation}

Now, as for $H[\xi]$, using the  translationally-invariant shift vector, $\xi^a = N^i \chi_i^a$, for constant $N^i$ yields

\begin{equation}
H(N^i) = \int d^3x \, p^{ab} \mathcal{L}_{N^i \chi_i^a} g_{ab} = -N^l k^{ij} q_{jk} \int d^3x \, \sqrt{\omega_0} \, \chi_i^a \chi_l^b \nabla_a \sigma^k_b.
\end{equation}
Noting that $\sigma^i$'s form an involutive distribution, as $d\sigma^i = \frac{1}{2} \epsilon^i_{\,jk} \sigma^j \wedge \sigma^k$, and given $\mathcal{L}_{\sigma} g_{ab} = 0$ due to translational isometry, $\nabla_{(a} \sigma^k_{\smash{b)}} = 0 $ and therefore, $\nabla_a \sigma_b^k = \frac{1}{2} d\sigma^k_{ab}$. This all results in

\begin{equation}\label{eq:cosmconstraintH}
H(N^i)=\frac{1}{2} \epsilon^k_{\,li} N^l k^{ij} q_{k j}.
\end{equation}
As a constraint for all $N^k$, we have the simple constraint on shape variables

\begin{equation}\label{eq:cosmconstraint2}
[\theta, \eta]_i^{\,j}=0.
\end{equation}

Only the definition of scale-invariant quantities remains:

\begin{equation}
\Sigma=\int d^3x \, \sqrt{G^s_{abcd} \pi^{ab} \pi^{cd} }=\sqrt{\mu_{ijkl} \eta^{ij} \eta^{kl}} .
\end{equation}
And as before,

\begin{equation}
\hat\eta^{i j}=\frac{\eta^{i j}}{\Sigma}, \quad\hat{\sigma}=\frac{\sigma^A}{\Sigma},\quad
\kappa=\frac{\Sigma^2}{v^\frac{4}{3}}, \quad\epsilon=\frac{v^2 \Lambda}{\Sigma^2},\quad \hat{H}_m=\Sigma H_m.
\end{equation}
The variables $\hat{\eta}^{i j}$ now satisfy an additional condition:

\begin{equation}\label{eq:cosmconstraint3}
\mu_{i j k l} \hat{\eta}^{ij} \hat\eta^{kl}=1.
\end{equation}

We shall now proceed to derive the equation of state. The decoupling condition $\frac{d}{d t}\sqrt{|g|}=c \sqrt{|g|}$ yields

\begin{equation}
N_0=\frac{2 c}{\sqrt{6 \left\lvert\frac{\mu_{ijkl} \eta^{ij} \eta^{kl}}{2 v^2}-v^{-\frac{2}{3}} R + 2 \Lambda+H_m\right\rvert}},
\end{equation}
and $\hat{N}_0=\frac{\Sigma}{v} N_0$ is the scale-independent lapse function.
Finally, the homogeneous system of equations can be derived from \cref{eq:cosmconstraintS}:

\begin{equation}\label{eq:cosmexpliciteqs}
\begin{array}{ccl}
d\theta_{i j} &=& \hat{N}_0 \mu_{i j k l} \hat{\eta}^{kl},\\
d\hat{\eta}^{i j}&=&\frac{\partial \hat{\eta}^{ij}}{\partial \theta_{k l}} d\theta_{k l} + \frac{\partial \hat{\eta}^{i j}}{\partial \eta^{kl}} d\eta^{kl}\\
&=& \frac{\partial \hat{\eta}^{ij}}{\partial \theta_{k l}} d\theta_{k l}
- (\delta^{(i}_k\delta^{j)}_l - \hat\eta^{ij} \mu_{klmn} \hat{\eta}^{mn}) \left(\frac{1}{2} \frac{\partial \mu_{mnop}}{\partial \theta_{kl}} \hat\eta^{mn} \hat\eta^{op}-\frac{1}{\kappa} \frac{\partial R}{\partial \theta_{kl}}\right),\\
d \kappa&=&-\frac{4}{3} \kappa c+2 \kappa \hat{N}_0 \mu_{i j k l} \hat{\eta}^{k l}\left(\frac{1}{\kappa} \frac{\partial R}{\partial \theta_{i j}}\right),\\
d \epsilon &=& 2 c \epsilon-2 \epsilon \hat{N}_0 \mu_{i j k l} \hat{\eta}^{k l}\left(\frac{1}{\kappa} \frac{\partial R}{\partial \theta_{i j}}\right),\\
d \phi_A &=& \kappa^{-\frac{3}{2}} \hat{N}_0 \frac{\partial \hat{H}_m} {\partial \hat{\sigma}^A},\\
d \hat{\sigma}^A &=& -\kappa^{-\frac{3}{2}} \hat{N}_0 \frac{\partial \hat{H}_m}{\partial \phi_A}-\hat{\sigma}^A \hat{N}_0 r_{i j k l} \hat{\eta}^{k l}\left(\frac{1}{\kappa} \frac{\partial R}{\partial \theta_{i j}}\right) .
\end{array}
\end{equation}
The variables have to satisfy the constraints in \cref{eq:cosmconstraints1,eq:cosmconstraint2,eq:cosmconstraint3}. Note that the symmetry-reduced system \cref{eq:cosmexpliciteqs} enjoys the same interpretation as the original system \cref{eq:expliciteqs}, with $\theta_{ij}$ and $\phi_A$ defining the associated shape space, $\hat\eta^{ij}$ and $\hat\sigma^A$ being the direction of change, and the additional variables $\kappa$ and $\epsilon$ being related to higher-order geometrical properties of the curve in shape space. This system, derived from full inhomogeneous PSD, can be used as the starting point for the analysis in \cite{Koslowski.2018}. Hence the evolution through the Big Bang, as derived in that paper, is a prediction of full inhomogeneous PSD.

\section{Conclusion}

As the second paper in the development of the classical PSD programme, following \cite{Koslowski.2022}, Einstein's GR has been studied and cast into the general framework of PSD.

Despite Einstein's initial aspirations, GR never originally developed into a truly relational theory, as the absolute scale is conspicuously part of the dynamical version in \cref{eq:ADMcons}. In the present paper, we have shown that the implementation of full spatial and time reparametrization relationalism is indeed a feature of GR by showing that the local conformal degrees of the spatial metric decouple, i.e., form an autonomous dynamical subsystem, from the scale and duration degrees of freedom on spacetime geometry (cf. Sec.~\ref{sec:decoupling}). In so doing, we have shown that the whole dynamical system of GR and matter fields can seamlessly construct a PSD system of variables in terms of only the geometrical properties of a curve in shape space, free of scale or any choice of external reference system or process. The system of equations \cref{eq:expliciteqs} defines the PSD description of Einstein's gravity in the most general setting on a compact Riemannian 3-dimensional manifold, for any matter component. Remarkably, despite the vast conceptual differences between the standard spacetime approach to GR and Newtonian particle mechanics, the PSD framework of relational mechanics exposes a remarkable similarity between the underlying fundamental structures of both theories, as explored in Sec.~\ref{subsec:comparison}.

As an application, we have derived the homogeneous Bianchi IX model from the PSD description of full GR. Interestingly, this analysis has shown that the Janus point evolution through the Big Bang, as found in \cite{Koslowski.2018}, is not an accident of the symmetry-reduced theory, but a generic feature of the full inhomogeneous PSD description of GR, and therefore, is expected to be a general result of the full PSD model.

Given the framework developed in this paper, three main questions and issues arise, that motivate future exploration:

1. Once all background structures that are normally used to define reference external and absolute systems and processes for measuring distance and time are eliminated from the fundamental theory, as in the present PSD description, the relational theory is required to develop stable rods and clocks w.r.t. which effective physics emerges. This problem of rods and clocks has been addressed in the simpler particle model (see \cite{Barbour.2014} for the seminal paper within standard SD, and \cite{Koslowski.2022} for the analogue analysis within PSD), but requires a rigorous exploration in the present field-theoretic model. 

2. Configuration complexity plays a central role in the Janus point approach to the problem of the arrow of time. The counterpart of configuration complexity, along with its triple role, namely i) the driving force in shape space, ii) a measure of structure formation, and iii) the defining property of the intrinsic arrow of time for all dynamical trajectories, is missing in this model (again, see \cite{Barbour.2014} for the particle model). Though clearly the Ricci scalar plays the role of shape potential in \cref{eq:expliciteqs}, a complexity function defined on the space of metric-shapes $\rho_{ab}$ and material fields, which quantifies structure formation and captures the dynamical generation of the arrow of time in keeping with the dynamics of shapes, is yet to be found. One natural candidate is the Shannon entropy of the matter field configuration, which will be the subject of a follow-up study. 

3. As mentioned in Sec.~\ref{sec:metricmeani}, a more radical implementation of the relational framework that underlies PSD suggests that the spatial conformal geometry itself is questionable as a ``fundamental'' field component of the universe. Therefore, the development of a description of PSD in terms of an equation of state of the geometry of the curve in the matter field sector of shape space is a very natural extension of the present framework.

\section*{Acknowledgements}

Tim Koslowski would like to thank Julian Barbour, Henrique Gomes, Sean Gryb, Flavio Mercati, David Sloan, Timothy Budd, Antonio Vassallo and Melissa Rodriguez for their collaboration on the Shape Dynamics programme. His foremost thanks go to Julian Barbour, who has developed many of the conceptual ideas exploited in this research field. Special thanks go also to Henrique Gomes, Flavio Mercati, David Sloan and Sean Gryb for many stimulating discussions about Pure Shape Dynamics. Additional thanks go to Daniel Sudarsky, Chryssomalis Chrysomalakos, Yuri Bonder, Marcelo Salgado and David Wiltshire for many stimulating and critical discussions that led to the PSD framework. Pedro Naranjo would also like to thank Julian Barbour for his hospitality at College Farm, where P. N.'s interests in relational physics started to develop, as well as for recent discussions. Also, discussions with Sean Gryb and Flavio Mercati are much appreciated. Finally, conceptual and philosophical discussions on relational physics with Antonio Vassallo are highly appreciated. Pooya Farokhi is deeply grateful to Julian Barbour for his support, encouragement, and many valuable discussions. P. F. appreciates the scholarship from Bonn-Cologne Graduate School which has facilitated the progress of this research. 

\newpage
\printbibliography
\end{document}